\documentclass[journal]{IEEEtran}

\ifCLASSINFOpdf
\else
   \usepackage[dvips]{graphicx}
\fi
\usepackage{url}

\usepackage{amsmath,amssymb,amsfonts}
\usepackage{algorithmic}
\usepackage{graphicx}
\usepackage{textcomp}
\usepackage[table]{xcolor}
\usepackage{multirow}
\usepackage{arydshln}
\usepackage{booktabs}
\usepackage{cite}

\begin{document}

\title{StreamMel: Real‑Time Zero-shot Text‑to‑Speech via Interleaved Continuous Autoregressive Modeling}

\author{Hui Wang, Yifan Yang, Shujie Liu, \IEEEmembership{Member, IEEE}, Jinyu Li, \IEEEmembership{Fellow, IEEE}, Lingwei Meng,\\ Yanqing Liu, Jiaming Zhou, Haoqin Sun, Yan Lu, and  Yong Qin \IEEEmembership{Member, IEEE}
\thanks{Hui Wang, Jiaming Zhou, Haoqin Sun, and Yong Qin are with the College of Computer Science, Nankai University, China. Yifan Yang, Shujie Liu, Jinyu Li, Lingwei Meng, Yanqing Liu, and Yan Lu are with the Microsoft Corporation. Shujie Liu, Jinyu Li, and Yong Qin are the corresponding authors (e-mail: shujliu@microsoft.com; jinyli@microsoft.com; qinyong@nankai.edu.cn). The work was done during the first author's internship at Microsoft.}
}

\markboth{Journal of \LaTeX\ Class Files, Vol. 14, No. 8, August 2015}
{Shell \MakeLowercase{\textit{et al.}}: Bare Demo of IEEEtran.cls for IEEE Journals}
\maketitle

\begin{abstract}
Recent advances in zero-shot text-to-speech (TTS) synthesis have achieved high-quality speech generation for unseen speakers, but most systems remain unsuitable for real-time applications because of their offline design. Current streaming TTS paradigms often rely on multi-stage pipelines and discrete representations, leading to increased computational cost and suboptimal system performance. In this work, we propose StreamMel, a pioneering single-stage streaming TTS framework that models continuous mel-spectrograms. By interleaving text tokens with acoustic frames, StreamMel enables low-latency, autoregressive synthesis while preserving high speaker similarity and naturalness. Experiments on LibriSpeech demonstrate that StreamMel outperforms existing streaming TTS baselines in both quality and latency. It even achieves performance comparable to offline systems while supporting efficient real-time generation, showcasing broad prospects for integration with real-time speech large language models. Audio samples are available at: \url{https://aka.ms/StreamMel}.
\end{abstract}


\IEEEpeerreviewmaketitle

\vspace{-5pt}
\section{Introduction}

\IEEEPARstart{Z}{ero-shot} Text-to-Speech (TTS) enables high-quality speech synthesis for unseen speakers using a few seconds of reference utterance. Recent advances in large language models (LLMs)~\cite{gpt4,audiolm} and generative methods~\cite{flowmatching,diffusion} have greatly improved both the expressiveness and the speaker consistency of TTS systems~\cite{valle, valle2, cosyvoice, seedtts, cosyvoice3, palle}. However, most existing zero-shot TTS systems are designed for offline scenarios, which require processing the entire input text before audio synthesis begins~\cite {valle, melle}. While effective in controlled settings, this approach is fundamentally misaligned with the demands of real-time applications, which require low latency and incremental synthesis as contextual information arrives in a streaming fashion. This mismatch limits the applicability of zero-shot TTS in latency-sensitive scenarios, such as simultaneous interpretation and live conversational agents. As a result, there is growing interest in developing streaming-capable zero-shot TTS systems that can meet the real-time requirements of these emerging use cases.

Streaming zero-shot TTS presents a unique set of challenges. A common limitation of existing streaming TTS systems lies in their complex two-stage modeling pipelines, which introduce latency bottlenecks. For example, hybrid approaches such as CosyVoice 2~\cite{cosyvoice2} and IST-LM~\cite{istlm} first employ a language model (LM) to predict semantic representations, followed by a separate acoustic flow-matching model to generate speech features. While the LM can operate token-by-token, the subsequent acoustic stage typically requires a buffer of accumulated tokens to produce stable outputs, thus increasing latency. Similarly, SMLLE~\cite{smlle} introduces an additional transducer module to better support streaming synthesis, but increases the architectural complexity of the system and introduces additional latency during real-time generation.

Another major limitation of these systems is their reliance on discrete representations, which fundamentally restricts the modeling capacity. Recent streaming zero-shot TTS models, such as SyncSpeech~\cite{syncspeech}, SMLLE~\cite{smlle}, and CosyVoice 2~\cite{cosyvoice2}, often rely on discrete representations. Although these representations are naturally compatible with language modeling frameworks, they require a quantization process that inevitably degrades representational fidelity and increases system complexity. These issues collectively impair synthesis fidelity and learning efficiency compared to continuous representations~\cite{puvvada2024discrete,ardit}. Similar limitations have also been observed in the field of computer vision~\cite{mar}. In contrast, recent TTS approaches based on continuous representations have demonstrated superior performance~\cite{melle, felle, ditar}, highlighting the advantages of preserving the rich structure inherent in continuous acoustic features.

To address these challenges, we propose \textbf{StreamMel}, the first single-stage streaming zero-shot TTS system based on continuous mel-spectrogram representations. Unlike prior approaches that rely on discrete quantized tokens, StreamMel directly models interleaved text and acoustic features in a unified, fully incremental framework. This design reduces system complexity, avoids information loss from quantization, and enables low-latency synthesis with high speech quality. Experiments on LibriSpeech demonstrate that StreamMel achieves lower latency than existing streaming zero-shot TTS models while maintaining speech quality comparable to offline systems. Our contributions are mainly threefold:

\begin{itemize}
    \item We propose a novel single-stage streaming zero-shot TTS framework that unifies text and acoustic modeling into a single autoregressive process, eliminating intermediate stages and significantly reducing latency.
    
    \item We introduce a continuous Mel interleaving strategy that seamlessly integrates phoneme and acoustic features in streaming, effectively preserving prosody and speaker characteristics by avoiding discretization.
    
    \item  StreamMel achieves state-of-the-art latency and speech quality among existing streaming zero-shot TTS systems. Our analysis further reveals key factors influencing the trade-off between latency and quality.
\end{itemize}

\section{StreamMel}

\begin{figure}
\centerline{\includegraphics[width=\columnwidth]{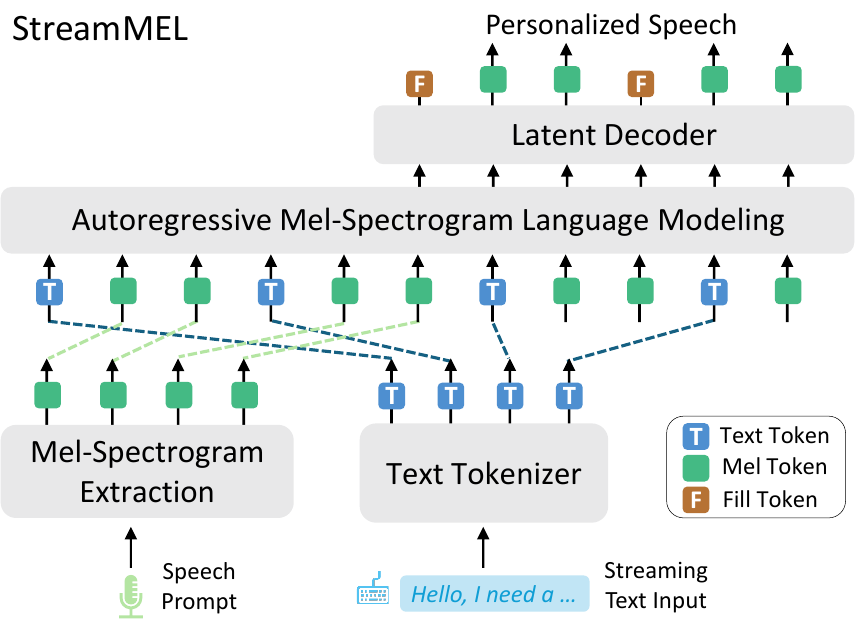}}
\vspace{-4pt}
\caption{StreamMel architecture. The model takes an interleaved sequence of text and mel-spectrogram tokens as input and autoregressively predicts the next token for streaming speech synthesis. Fill tokens mark positions without prediction targets and are ignored during loss computation.}

\label{fig:overview}
\vspace{-8pt}
\end{figure}

\subsection{Problem Formulation}

We formulate streaming zero-shot TTS as an autoregressive generation task over an interleaved sequence of phonemes and continuous acoustic tokens. Given a text prompt encoded as a sequence of phonemes $\mathbf{x} = [x_0, x_1, \dots, x_{L-1}]$, and a corresponding mel-spectrogram $\mathbf{y} = [y_0, y_1, \dots, y_{T-1}]$, we define a joint input sequence $\mathbf{z}$ through an interleaving process. Specifically, let $n$ denote the number of phoneme tokens and $m$ the number of mel-spectrogram frames between consecutive phoneme segments. The interleaved sequence is constructed as
\begin{align}
\mathbf{z} = [&x_0, \dots, x_{n-1}, y_0, \dots, y_{m-1}, \nonumber \\
             &x_n, \dots, x_{2n-1}, y_m, \dots, y_{2m-1}, \dots].
\label{eq:interleave}
\end{align}
This interleaving continues at a fixed $n\!:\!m$ ratio, followed by all the remaining mel-spectrogram frames once phonemes are exhausted.
Each mel frame $y_t$ is generated autoregressively, conditioned on all previous tokens in the interleaved sequence $\mathbf{z}_{<t'}$, where $t'$, the position of $y_t$ in $\mathbf{z}$, is computed as:
\begin{equation}
t' =
\begin{cases}
t + \left\lfloor \dfrac{t}{m} \right\rfloor \cdot n, & \text{if } \left\lfloor \dfrac{t}{m} \right\rfloor \cdot n < L \\
t + L, & \text{otherwise}
\end{cases}
\label{eq:position}
\end{equation}

The model autoregressively estimates the conditional distribution over mel frames as:
\begin{equation}
p(\mathbf{y} \mid \mathbf{x}; \theta) = \prod_{t=0}^{T - 1} p(y_t \mid \mathbf{z}_{<t'}; \theta).
\label{eq:full-prob}
\vspace{-5pt}
\end{equation}

\subsection{Overview}

As shown in Fig.~\ref{fig:overview}, the proposed StreamMel framework consists of three major components: an interleaved input strategy that combines linguistic and acoustic tokens, an autoregressive mel-spectrogram language model, and a lightweight latent decoder module. This design enables fully streaming zero-shot speech synthesis with low latency, continuous representation modeling, and expressive output quality.

\subsubsection{Interleaved Input}

To enable streaming synthesis, the input modalities are processed separately: the text prompt is tokenized using a text tokenizer, while the speech prompt is converted into continuous mel-spectrogram frames and projected into mel tokens via a pre-net, which is a multi-layer perceptron (MLP). These tokens are interleaved to form a unified input sequence. Crucially, new text tokens can be appended incrementally, allowing the generation to begin without requiring full utterance in advance. At each decoding step, the model autoregressively predicts the next mel-spectrogram frame in the interleaved sequence $\mathbf{z}$, while the text tokens are streamingly copied from the coming text input, rather than predicted, arranged as in Equation~\eqref{eq:interleave}. This design supports causal inference and allows fine-grained control over the balance between the linguistic and acoustic context.

\subsubsection{Mel-spectrogram Language Model}

The core of StreamMel is an autoregressive language model operating over a unified sequence of text and mel-spectrogram representations. 
The model is implemented as a Transformer decoder with causal masking, enabling frame-synchronous, left-to-right generation. This architecture allows the model to incrementally incorporate newly arrived text tokens while conditioning on the full history of previously generated acoustic frames.

\subsubsection{Latent Decoder}
To improve the expressiveness and controllability of speech generation, we introduce a two-part output module comprising latent sampling and stop prediction. The latent sampling module enhances the diversity and naturalness of generated speech, following the approach proposed in MELLE~\cite{melle}. At each decoding step $t'$, if the current position corresponds to a non-fill token (i.e., a mel-spectrogram frame), the decoder output $e_{t'}$ is used to predict a mean $\boldsymbol{\mu}_{t'}$ and a log-variance $\log \boldsymbol{\sigma}_{t'}^2$ of a gaussian distribution $\mathcal{N}(\mu_{t'}, \sigma_{t'}^2)$. A latent variable is then sampled using the reparameterization trick as $\mathbf{z}_{t'} = \boldsymbol{\mu}_{t'} + \boldsymbol{\sigma}_{t'} \odot \boldsymbol{\epsilon}, \quad \boldsymbol{\epsilon} \sim \mathcal{N}(0, \mathbf{I}),$
where $\odot$ denotes element-wise multiplication. The latent vector $\mathbf{z}_{t'}$ is projected to the mel-spectrogram space via a lightweight MLP to generate the output frame $\hat{\mathbf{y}}_t$. No output is produced at fill token positions. In parallel, the stop prediction module estimates a binary stop probability directly from $e_{t'}$, enabling the model to determine when to terminate the generation process. This mechanism allows for flexible-length inference.

\subsection{Loss Functions}
We adopt a standard multi-task objective, consisting of a mel-spectrogram regression loss, a Kullback-Leibler (KL) divergence loss~\cite{kingma2013auto, melle}, a spectrogram flux loss for dynamic variation~\cite{melle}, and a stop prediction loss~\cite{tacotron}:

\begin{equation}
\mathcal{L} = \alpha \mathcal{L}_{\text{reg}} + \lambda \mathcal{L}_{\text{KL}} + \beta \mathcal{L}_{\text{flux}} + \gamma \mathcal{L}_{\text{stop}}.
\label{eq:loss}
\end{equation}
The regression loss $\mathcal{L}_{\text{reg}}$ is a combination of L1 and L2 distances between the predicted and ground-truth mel spectrogram, ensuring accurate spectral reconstruction. The KL divergence loss $\mathcal{L}_{\text{KL}}$ regularizes the latent variable distribution towards a standard Gaussian prior. The spectrogram flux loss $\mathcal{L}_{\text{flux}}$ encourages expressive temporal variations by penalizing discrepancies in the frame-to-frame difference. Finally, the stop prediction loss $\mathcal{L}_{\text{stop}}$ is a binary cross-entropy loss that guides the model to accurately predict the end of synthesis.

\section{Experiment}

\begin{table}
\centering
\renewcommand{\arraystretch}{1.05}
\caption{Continuation objective performance between StreamMel and baseline models. \textbf{Bold} indicates the best streaming model. GT denotes ground truth. MELLE (L) refers to the MELLE-limited variant as defined in~\cite{melle}.
}
\label{tab:cont}
\begin{tabular}{l cccc}
\toprule
\textbf{Model} & \textbf{WER-C} & \textbf{WER-H} & \textbf{SIM-R} & \textbf{SIM-O} \\
\midrule
GT & 1.41 & 1.91 & - & 0.668 \\
\rowcolor{gray!20}\multicolumn{5}{l}{\textbf{Non-streaming}} \\
VALL-E R & 1.58 & 2.32 & 0.397 & 0.363 \\
MELLE (L) & 1.53 & 2.22 & 0.517 & 0.480 \\
FELLE & 1.53 & 2.27 & 0.539 & 0.513 \\
\rowcolor{gray!20}\multicolumn{5}{l}{\textbf{Streaming}} \\
IST-LM & - & 3.60 & - & - \\
StreamMel & \textbf{1.65} & \textbf{2.41} & \textbf{0.534} & \textbf{0.504} \\
\bottomrule
\end{tabular}

\end{table}

\begin{table}
\centering\renewcommand{\arraystretch}{1.05}
\caption{Cross-sentence subjective performance between StreamMel and baseline models.}
\label{tab:sub}
\begin{tabular}{l cccc}
\toprule
& \textbf{GT} & \textbf{SMLLE} & \textbf{IST-LM} & \textbf{StreamMel} \\
\midrule
\textbf{MOS} & $4.32_{\pm0.15}$ & $3.10_{\pm0.19}$ & $4.12_{\pm0.16}$ & \textbf{4.14$_{\pm0.16}$} \\
\textbf{SMOS} & $4.41_{\pm0.14}$ & $3.36_{\pm0.17}$ & $4.22_{\pm0.14}$ & \textbf{4.27$_{\pm0.15}$} \\
\bottomrule
\end{tabular}
\end{table}

\subsection{Implementation Details}

We perform experiments using the LibriSpeech dataset~\cite{librispeech}, which consists of around 960 hours of English read speech. Zero-shot TTS evaluation is performed on the Librispeech test-clean set, ensuring that test speakers are excluded from the training set. Following recent protocols~\cite{valler, melle}, we sample utterances of 4 to 10 seconds in duration for evaluation.

Our model follows the Transformer-based architecture~\cite{valle, melle}, with 12 Transformer decoder blocks, each having 16 attention heads and a feed-forward layer of dimension 4,096. The hyperparameters are set as: $\alpha = 2$, $\lambda = 0.05$, $\beta = 1$, and $ \gamma = 0.5$, balancing the contributions of each loss component in the overall objective function. During training and evaluation, we adopt a fixed interleaving ratio of $1\!:\!4$.

\subsection{Evaluation Setting}

We evaluate our model on two scenarios for zero‐shot TTS: \textbf{continuation}, where the model receives a transcript and the first three seconds of an utterance from the target speaker and is expected to continue the speech in the same voice and style; and \textbf{cross‐sentence}, where the model is given a reference utterance and its transcription from a speaker along with a new target sentence, and must synthesize the target sentence using the voice characteristics of the reference speaker.

We evaluate our model against baseline systems, including non-streaming models such as VALL-E R and MELLE, which use full-context inputs for high-quality synthesis~\cite{f5tts,maskgct,cosyvoice,cosyvoice2,valle}, and streaming models such as IST-LM and SMLLE, which are designed for low-latency, real-time speech generation. These baselines provide comprehensive coverage of both high-fidelity and streaming synthesis paradigms.
\vspace{-2pt}

\subsection{Evaluation Metric}

We use four metrics to evaluate system performance: (1) \textbf{Speech quality} is assessed via crowd-sourced Mean Opinion Score (MOS) which is widely regarded as the gold standard~\cite{ramp,ramp+}; Speaker Mean Opinion Score (SMOS) is used to evaluate the similarity between the synthesized and target speaker voices. (2) \textbf{Intelligibility} is measured by Word Error Rate (WER) using Conformer-Transducer~\cite{conformer}, HuBERT-Large~\cite{hubert}, and Whisper-Large-V3~\cite{whisper}, denoted as WER-C, WER-H, and WER-W; (3) \textbf{Speaker similarity (SIM)} is computed via cosine similarity between speaker embeddings from WavLM-TDNN~\cite{wavlm}, with SIM-O (vs. original audio) and SIM-R (vs. vocoder-reconstructed audio); (4) \textbf{Streaming performance} includes First-Packet Latency (FPL), defined as the time from receiving text input to the start of speech generation, evaluated under FPL-A (immediate text) and FPL-L (with upstream LLM delay)~\cite{syncspeech}; and {Real-Time Factor (RTF)}, the ratio of synthesis time to output audio duration.

\begin{table}
  \centering
  \renewcommand{\arraystretch}{1.2}
  \caption{Cross-sentence objective performance between StreamMel and baseline models under different training data scales.}
\vspace{-2pt}
  \label{tab:cross1}
  \resizebox{\columnwidth}{!}{%
    \begin{tabular}{llrrrr}
      \toprule
      \textbf{Model} & \textbf{Training Data} & \textbf{WER-C} & \textbf{WER-H} & \textbf{SIM-R} & \textbf{SIM-O} \\
      \midrule
      GT & – & 1.41  & 1.91  & – & 0.778 \\
      \rowcolor{gray!20}\multicolumn{6}{l}{\textbf{Non-streaming}} \\
      VALLE-R          & LibriSpeech  & 3.18  & 3.97  & 0.395 & 0.365 \\
      
      MELLE (L)   & LibriSpeech  & 2.21  & 2.80  & 0.633 & 0.591 \\
      FELLE  & LibriSpeech  & 2.20  & 2.89  & 0.654 & 0.619 \\
      MELLE            & Libriheavy   & \underline{1.47}  & \underline{2.10}  & 0.664 & 0.625 \\
      VALL-E 2         & Libriheavy  & 1.50  & 2.44  & \underline{0.678} & \underline{0.642} \\
      \rowcolor{gray!20}\multicolumn{6}{l}{\textbf{Streaming}} \\
      SMLLE         & LibriSpeech  & 5.14  & 6.37  & 0.516 & 0.489 \\
      IST-LM     & LibriTTS   & –     & 4.53  & –     & \textbf{0.653} \\
      StreamMel        & LibriSpeech  & \textbf{2.10}  & \textbf{2.76}  & \textbf{0.656} & 0.622 \\
      \bottomrule
    \end{tabular}%
  }

\end{table}

\vspace{-2pt}
\section{Result}

\subsection{Comparative Result}

\paragraph{Continuation Performance}  
As shown in Table~\ref{tab:cont}, StreamMel achieves a competitive trade-off between speech quality and speaker similarity for the continuation task. Compared to non-streaming models like MELLE and VALL-E R, StreamMel slightly increases WER-C but outperforms them on similarity, indicating improved speaker consistency. Compared to IST-LM~\cite{istlm}, another streaming baseline, StreamMel demonstrates significantly better WER-H, highlighting the effectiveness of continuous representations in improving both intelligibility and speaker fidelity in streaming TTS.

\begin{table}
  \centering
  \renewcommand{\arraystretch}{1.1}
  \caption{Cross-sentence objective performance between StreamMel and baseline models. Results marked with $^{*}$ are from~\cite{syncspeech}.}
  \label{tab:latency}

    \begin{tabular}{llrrrr}
      \toprule
      \textbf{Model} & \textbf{Training Data} & \textbf{WER-W} & \textbf{FPL-A} (s) & \textbf{FPL-L} (s) \\
      \midrule
      GT             & –             & 1.70   & –    & –     \\
      \rowcolor{gray!20}\multicolumn{6}{l}{\textbf{Non-streaming}} \\
      VALL-E$^{*}$        & Libriheavy   & 5.90 & 0.75  & 1.47  \\
      MASKGCT$^{*}$   & Emilia    & 2.77   & 2.15  & 2.55  \\
      CosyVoice$^{*}$     & LibriTTS  & 3.47   & 0.22  & 0.94  \\
      CosyVoice 2$^{*}$   & LibriTTS   & 3.00   & 0.22  & 0.35  \\
      F5-TTS$^{*}$        & Emilia   & 2.51 & 1.27  & 1.98  \\
      \rowcolor{gray!20}\multicolumn{6}{l}{\textbf{Streaming}} \\
      SyncSpeech$^{*}$    & LibriTTS     & 3.07  & 0.06  & 0.11  \\
      StreamMel      & LibriSpeech & \textbf{2.77}  & \textbf{0.01}  & \textbf{0.04}  \\
      \bottomrule
    \end{tabular}
  
\end{table}

\begin{table}
  \centering
  \caption{Cross-sentence objective performance of StreamMel under different $n\!:\!m$ ratios.}
  \label{tab:gt_streammel_nm}
  \begin{tabular}{lccccc}
    \toprule
    $n\!:\!m$ & \textbf{WER-C} & \textbf{WER-H} & \textbf{WER-W} & \textbf{SIM-R} & \textbf{SIM-O} \\
    \midrule
   GT   & 1.41 & 1.91 & 1.70 & -     & 0.778 \\
    \midrule
   Non-streaming & 1.95 & 2.42 & 2.37 & 0.643 & 0.606 \\
   3:1 & 3.55 & 4.38 & 4.26 & 0.649 & 0.610 \\
   2:1 & 2.25 & 2.96 & 2.73 & 0.642 & 0.605 \\
   1:1 & 2.24 & 2.93 & 2.83 & 0.645 & 0.605 \\
   1:2 & 2.50 & 3.22 & 2.89 & 0.651 & 0.615 \\
   1:3 & 2.40 & 3.13 & 2.91 & 0.640 & 0.605 \\
   1:4 & 2.10 & 2.76 & 2.77 & 0.656 & 0.622 \\
   1:5 & 8.44  & 9.04   & 8.98  & 0.634  & 0.595 \\
    \bottomrule
  \end{tabular}
  
\end{table}

\paragraph{Cross Performance}  
From Table~\ref{tab:sub}, we observe that  StreamMel demonstrates clear advantages in both naturalness and speaker similarity over existing baselines. Compared to SMLLE and IST-LM, StreamMel produces speech that is consistently rated closer to the ground truth, suggesting that it can generate more natural and speaker-consistent audio.

Table~\ref{tab:cross1} demonstrates that StreamMel consistently leads in performance across most evaluation criteria among streaming models. When compared to non-streaming models trained on the same LibriSpeech dataset, such as MELLE~(L), StreamMel shows comparable or better speaker similarity and maintains competitive WER performance. Furthermore, although models like MELLE and VALL-E~2 benefit from training on larger datasets~\cite{libriheavy}, StreamMel still achieves similar SIM-R and SIM-O scores, suggesting its strong efficiency and competitiveness under more constrained training conditions.

Table~\ref{tab:latency} reports a WER-W of 2.77 for StreamMel, outperforming the streaming baseline and several non-streaming models. Although F5-TTS has a lower WER, it relies on non-streaming generation and a larger corpus~\cite{emilia}. In terms of latency, StreamMel shows the lowest first-packet latency (FPL). To generate the first audio packet, only $n$ TTS forward passes are required. When LLM latency is included, an additional $n$ LLM forward steps are needed. Thus, the first-packet latency is $L_{\text{FPL-L}}^{\text{StreamMel}} = d_{\text{LLM}} + d_{\text{TTS}}$ and $L_{\text{FPL-A}}^{\text{StreamMel}} = d_{\text{TTS}}$ for $n = 1$, where $d_{\text{LLM}} = 25\,\text{ms}$, based on Qwen-7B~\cite{syncspeech}, and $d_{\text{TTS}}$ is measured on a single NVIDIA A100 GPU.

\vspace{-2pt}
\subsection{Ablation Study}

\begin{table}
  \centering
  \caption{Cross-sentence objective performance of StreamMel under different reduction factors $r$.}
  \label{tab:reduction}
  \begin{tabular}{lcccccc}
    \toprule
    $r$ & \textbf{WER-C} & \textbf{WER-H} & \textbf{WER-W} & \textbf{SIM-R} & \textbf{SIM-O} & \textbf{RTF} \\
    \midrule
       GT & 1.41 & 1.91 & 1.70 & - & 0.778 & - \\
    \midrule
       1 & 2.24 & 2.93 & 2.83 & 0.645 & 0.605 & 0.700 \\
       2 & 2.34 & 3.99 & 2.77 & 0.598 & 0.552 & 0.352 \\
       3 & 2.17 & 2.97 & 2.47 & 0.543 & 0.492 & 0.233 \\
       4 & 1.96 & 2.80 & 2.32 & 0.512 & 0.454 & 0.179 \\
    \bottomrule
  \end{tabular}
  
\end{table}

\begin{figure}
\centerline{\includegraphics[width=\columnwidth]{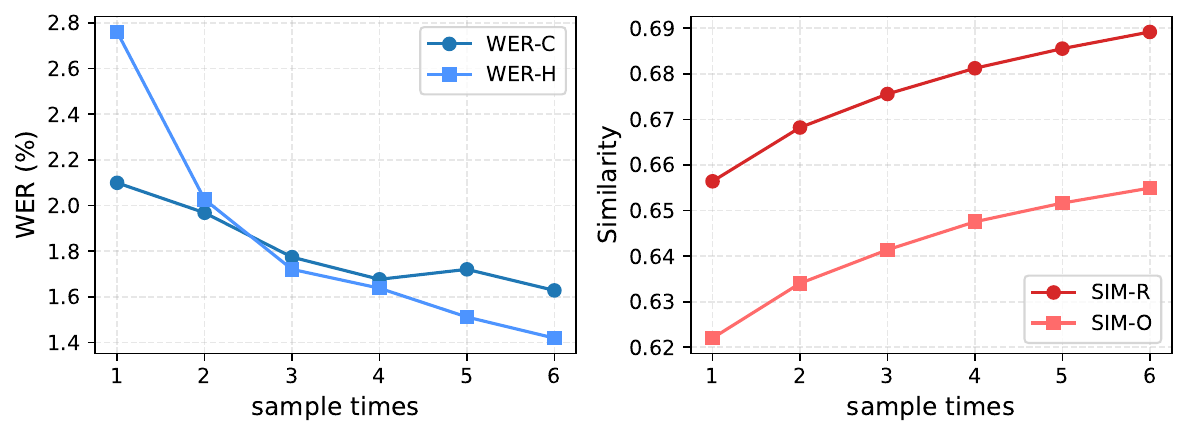}}
\vspace{-4pt}
\caption{Effect of sample times on WER (left) and similarity metrics (right).}
\label{fig:samples}
\end{figure}

\paragraph{Effect of Interleaving Ratio}  
We study how different interleaving ratios $n\!:\!m$ between text and mel-spectrogram tokens affect performance. Compared to the non-streaming setting, interleaving generally results in increased WER due to the additional complexity of joint linguistic and acoustic modeling. In addition, speech quality, measured by SIM-R and SIM-O, remains stable or slightly improves with moderate interleaving. Among all ratios, 1:4 yields the best balance. This suggests that sparse text token insertion provides sufficient semantic guidance without destabilizing acoustic prediction. In contrast, extreme ratios such as 3:1 and 1:5 significantly degrade both accuracy and quality.

\paragraph{Effect of Reduction Factor}  
We evaluates the impact of the reduction factor  $r$ under the interleaving setting $n\!:\!m = 1\!:\!1$, as reported in Table~\ref{tab:reduction}. Increasing $r$ leads to a consistent reduction in real-time factor (RTF), indicating improved inference efficiency. At the same time, WER metrics exhibit a general downward trend, suggesting that generating multiple frames per decoding step can mitigate error accumulation. In contrast, similarity metrics decline with larger $r$, reflecting a loss in acoustic detail and naturalness. These results illustrate the trade-off between decoding speed and output fidelity introduced by the reduction factor.

\paragraph{Effect of sample times} Sample times refers to how many latent variables are drawn during inference. In Fig.~\ref{fig:samples}, increasing sample times leads to a clear reduction in WER. At the same time, similarity scores increase steadily, reflecting increased consistency and naturalness in the generated speech. These results demonstrate that sampling times contribute to both clearer articulation and more natural-sounding output, with gains gradually saturating at higher sample counts.

\section{Conclusion}
We present StreamMel, the first single-stage streaming TTS framework that achieves efficient zero-shot speech synthesis by directly modeling interleaved sequences of text and continuous mel-spectrogram tokens. Experimental results confirm that StreamMel achieves state-of-the-art performance in both speech quality and latency compared to existing streaming TTS baselines. Looking ahead, StreamMel offers a compelling foundation for integration with real-time speech LLMs through continuous-valued representations.

\bibliographystyle{IEEEtran}
\bibliography{ref}

\end{document}